\documentclass[conference]{IEEEtran}
\IEEEoverridecommandlockouts
\usepackage{cite}
\usepackage{amsmath,amssymb,amsfonts}
\usepackage{algorithmic}
\usepackage{graphicx}
\usepackage{textcomp}
\usepackage{xcolor}
\usepackage{url}
\usepackage{eso-pic}
\usepackage{subcaption}
\def\BibTeX{{\rm B\kern-.05em{\sc i\kern-.025em b}\kern-.08em
    T\kern-.1667em\lower.7ex\hbox{E}\kern-.125emX}}

\newcommand{\ieeearxivnotice}{%
  \begingroup
  \scriptsize
  \setlength{\parindent}{0pt}%
  \begin{minipage}{1.0\textwidth}
  \textcopyright~2026 IEEE.
  Personal use of this material is permitted.
  Permission from IEEE must be obtained for all other uses,
  in any current or future media, including reprinting/republishing
  this material for advertising or promotional purposes,
  creating new collective works, for resale or redistribution to servers
  or lists, or reuse of any copyrighted component of this work in other works.
  DOI: 10.23919/NICOIntCPS00076.2026.00012.
  \end{minipage}
  \endgroup
}

\newcommand{\addieeearxivnotice}{%
  \AddToShipoutPictureFG*{%
    \AtPageLowerLeft{%
      \hspace{0.08\paperwidth}%
      \raisebox{0.45in}{\ieeearxivnotice}%
    }%
  }%
}

\makeatletter
\def\@IEEEBIOphotowidth{1.0in}
\def\@IEEEBIOphotodepth{1.15in}
\makeatother

\begin{document}

\title{Smooth Motion Stitching via \\Laplacian Optimization
in Rodrigues Vector Space
}

\author{\IEEEauthorblockN{1\textsuperscript{st} Ryosuke Higasayama}
\IEEEauthorblockA{\textit{Graduate School of Engineering} \\
\textit{Takushoku University}\\
Tokyo, Japan \\
25m310@st.takushoku-u.ac.jp}
\and
\IEEEauthorblockN{2\textsuperscript{nd} Hideki Todo}
\IEEEauthorblockA{\textit{Faculty of Engineering} \\
\textit{Takushoku University}\\
Tokyo, Japan \\
htodo@cs.takushoku-u.ac.jp}
\and
\IEEEauthorblockN{3\textsuperscript{rd} Jongseong Gwak}
\IEEEauthorblockA{\textit{Faculty of Engineering} \\
\textit{Takushoku University}\\
Tokyo, Japan \\
j-gwak@cs.takushoku-u.ac.jp}
}

\addieeearxivnotice
\maketitle

\newcommand{\fref}[1]{Figure~\ref{#1}}
\newcommand{\tref}[1]{Table~\ref{#1}}
\newcommand{\secref}[1]{Section~\ref{#1}}
\renewcommand{\vec}[1]{\mathbf{#1}}
\newcommand{\mat}[1]{\mathbf{#1}}

\begin{abstract}
This paper presents a motion editing framework for smooth motion stitching based on Laplacian optimization in Rodrigues vector space. By representing joint rotations as continuous Rodrigues vectors, motion stitching is formulated as a temporal Laplacian optimization problem, enabling smooth transitions between motion segments while preserving characteristic temporal variations of reference motions. The proposed approach supports both intra-category replacement and cross-category motion stitching without relying on learning-based models or complex manual tuning, and is computationally efficient for interactive editing. Through a series of stitching experiments and comparisons with linear interpolation, we demonstrate that Laplacian editing produces stable and visually coherent transitions under a wide range of motion differences. Furthermore, an analysis of rotational continuity clarifies that rotation-axis inversions are rare in real motion data and explains why numerical instabilities observed in synthetic axis-flipping scenarios do not arise in practical motion stitching. These results highlight the importance of rotational representation in stabilizing temporal optimization and suggest that the proposed framework is well suited not only for animation authoring but also for motion analysis and future extensions incorporating perceptual or physiological cues.
\end{abstract}


\section{Introduction}

Recent character animation workflows increasingly rely on combining multiple
motion capture sequences, making partial replacement and stitching common
operations in practical production pipelines.

However, temporal cut-and-paste of different motions often introduces
discontinuities in joint rotations, especially when stitched motions differ in
speed or rhythm, and simple interpolation results in unnatural transitions.

To address this issue, we propose a motion editing framework for correcting
rotational discontinuities at motion connections.
Joint rotations are represented using Rodrigues vectors, which enables the
application of temporal Laplacian editing to joint rotation sequences, allowing
boundary artifacts to be suppressed while preserving characteristic temporal
variations.

Through experimental analysis, we observe that rotation axis flipping is rare
in real motion capture data and at motion connections.
To examine the impact of axis inconsistencies, we conduct experiments with
intentionally inverted Rodrigues vector representations corresponding to the
same physical rotations, which introduce visible artifacts at stitching
boundaries.
In contrast, for real motion data in which axis flipping is detected, motion
segments can be stitched smoothly without noticeable artifacts using the
proposed formulation.

Experimental results on a variety of motion sequences demonstrate that the
proposed method enables stable and natural motion stitching across motions with
different temporal characteristics.

\section{Related Work}

\subsection{Stylized and Exaggerated Motion}

Numerous techniques have been proposed to exaggerate or stylize character
motions to enhance visual expressiveness.
Early work emphasized characteristic motion features through trajectory and
temporal modification, such as cartoon animation filters~\cite{wang2006cartoon}
and automatic expressive deformations~\cite{noble2006automatic}.
More recent approaches explore user-controllable motion exaggeration for
artistic direction, including artist-guided exaggeration methods such as
\emph{Smear}~\cite{basset2024smear}.

While these methods are effective for enhancing visual impressions, they
typically assume temporally continuous motion sequences and do not explicitly
address the problem of stitching independently captured motion segments with
smooth transitions.

\subsection{Motion Editing and Motion Stitching}

Motion editing through segmentation, recombination, and reuse of existing motion
data has been widely studied.
Representative examples include part-wise motion assembly~\cite{jang2022motion},
pattern-based motion synthesis~\cite{MotionTexture}, and learning-based
transition generation~\cite{li2023example}.

Several studies explicitly focus on transition control and motion
concatenation.
Physically motivated approaches enforce transition consistency using dynamic
constraints~\cite{Hubert}, while parametric motion graphs organize motion data
to define feasible transition paths~\cite{Rachel}.

In contrast to methods that rely on explicitly designed transition rules,
discrete graph structures, or parameterized connectivity, our approach adopts a
continuous optimization framework that directly smooths motion representations
along the temporal axis, enabling natural transitions without explicit
transition design.

Recent learning-based approaches, including keyframe-conditioned motion
in-betweening~\cite{Ren2024}, have demonstrated strong capabilities in generating
plausible transitions between sparse motion inputs.
These methods typically formulate motion stitching as a data-driven synthesis
problem, where intermediate frames are generated based on learned priors.

In contrast, our approach targets an editing-oriented scenario, where existing
motion segments are directly manipulated while preserving their original
temporal characteristics.
Rather than synthesizing new motion, we explicitly optimize the given motion
sequence to ensure continuity under user-defined edits, while minimizing
deviation from the original motion.
This provides deterministic control, reproducibility, and compatibility with
existing production pipelines, which are often difficult to guarantee in
generative approaches.

Moreover, our formulation enables direct optimization in a continuous
representation space, making it suitable for integrating temporal smoothness
constraints such as Laplacian regularization.
This contrasts with learning-based approaches, where such constraints are
implicitly handled by the model and are not directly controllable.

Therefore, our method is complementary to learning-based techniques:
while generative models are suitable for producing plausible transitions from
sparse inputs, our approach is particularly effective in scenarios requiring
precise editing of captured motion data, such as post-production adjustment,
motion refinement, and consistency preservation across stitched segments.

\subsection{Motion Representation and Optimization}

Motion editing performance strongly depends on the choice of rotational
representation.
Prior work has explored Euler angles, quaternions, and exponential maps, each
with different trade-offs in continuity and stability~\cite{tak2005physically,
hsu2005style}.
Laplacian-based editing has been widely used in geometry processing and has also
been applied to motion smoothing.

However, the interaction between rotational representations and temporal
Laplacian optimization has not been sufficiently examined in the context of
motion stitching.
In this work, we represent joint rotations using Rodrigues vectors and apply
Laplacian optimization directly to their temporal sequences to suppress
discontinuities at motion boundaries.
This distinguishes our approach from prior motion editing methods that rely on
discrete transition design, graph-based connectivity, or purely geometric
interpolation.

\section{Proposed Motion Stitching Framework}

\begin{figure*}[tb]
\centering
\includegraphics[width=1.0\linewidth]{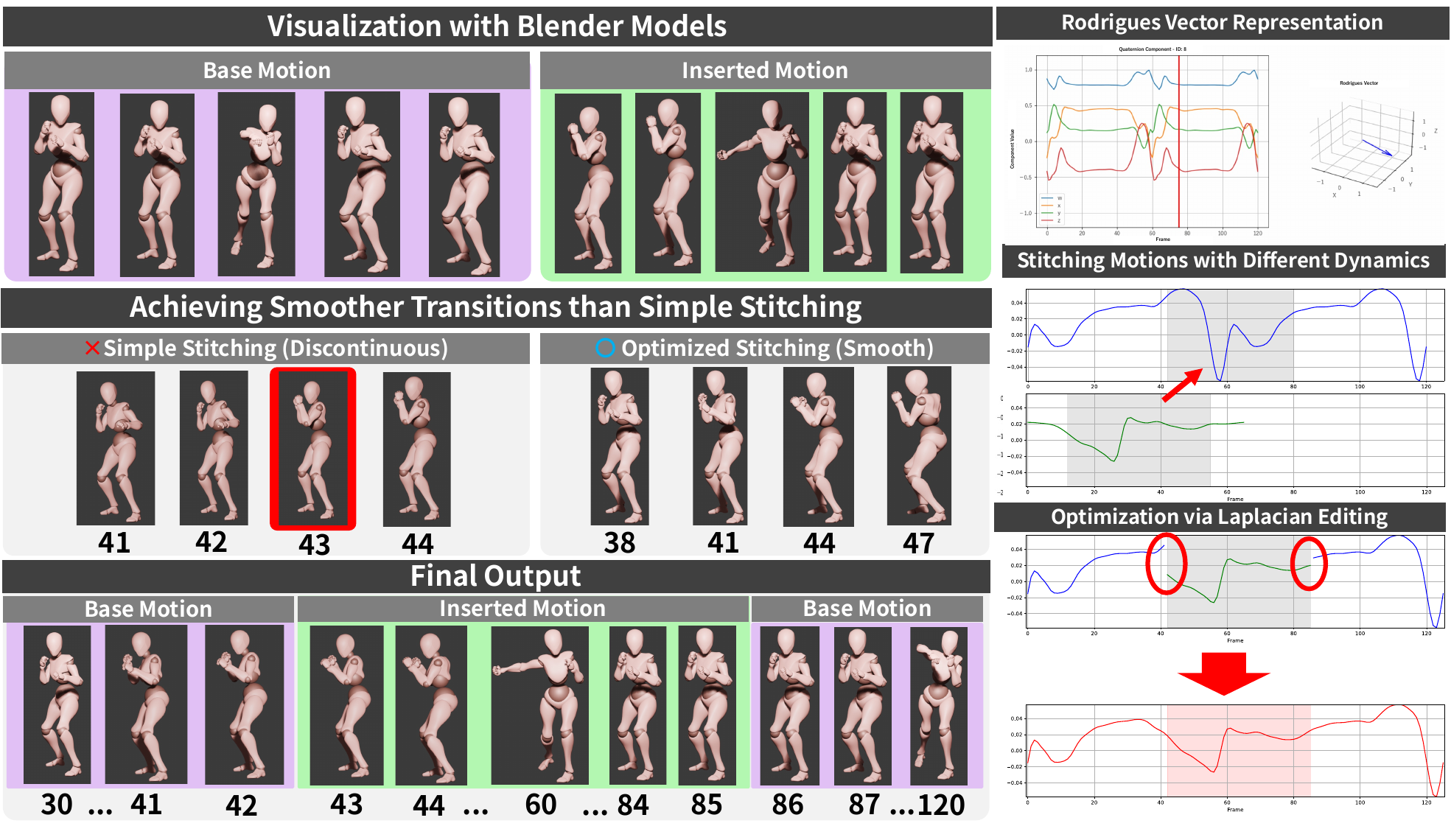}
\caption{
Overview of the proposed motion stitching framework, where smooth transitions
are achieved by Laplacian optimization in Rodrigues vector space.
}
\label{fig:overview}
\end{figure*}

\fref{fig:overview} provides an overview of the proposed motion stitching
framework.
Standard motion capture data are converted into Rodrigues vector
representations, which enables the application of temporal Laplacian editing to
joint rotation sequences.
Discontinuities introduced by cut-and-paste motion editing are suppressed
through Laplacian optimization, resulting in smooth and visually coherent
transitions without relying on learning-based models or complex parameter
tuning.

Based on this representation, users can select, replace, or insert motion
segments from reference motions.
Auxiliary temporal variation measures in Rodrigues vector space are computed for
analysis and qualitative evaluation, while the optimization framework focuses on
correcting rotational artifacts at motion boundaries.
As a result, motion segments with different temporal characteristics can be
integrated seamlessly within a unified framework.

\subsection{Conversion to Rodrigues Vector Representation}

\begin{figure}[tb]
\centering
\includegraphics[width=1.0\linewidth]{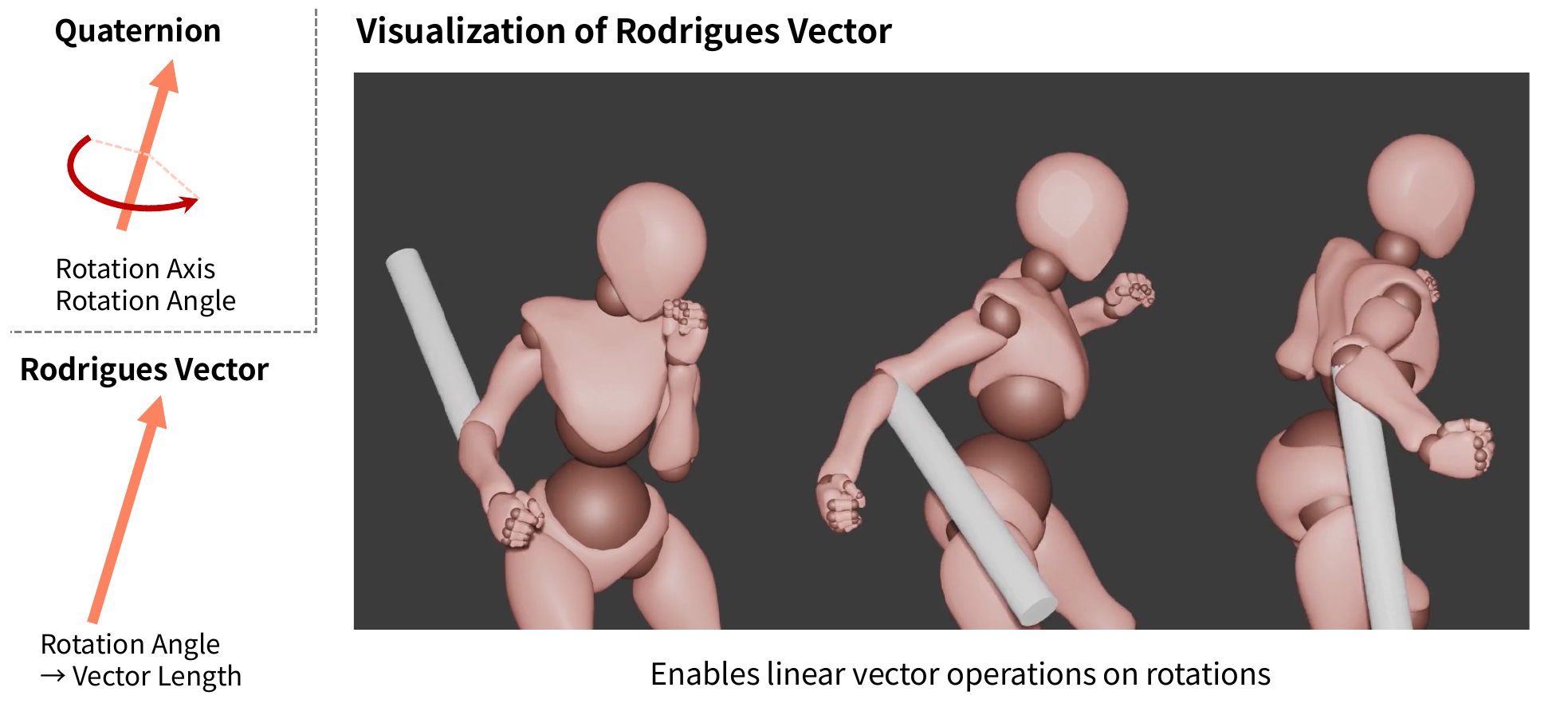}
\caption{Conversion from a quaternion representation to a Rodrigues vector.}
\label{fig:rodrigues_vector}
\end{figure}

In this work, joint rotations are represented using Rodrigues vectors
(\fref{fig:rodrigues_vector}).
A Rodrigues vector is defined as the product of a rotation angle $\theta$ and a
unit rotation axis $\vec{u}$, allowing joint rotations to be expressed as
three-dimensional vectors in Euclidean space.
This representation enables the direct application of Laplacian editing to
rotational motion.

Joint rotations stored as rotation matrices or quaternions are converted into
Rodrigues vectors so that each joint motion can be treated as a vector-valued
time series.
Given a quaternion $q = (w, \vec{v})$, the conversion is performed as follows:
\begin{equation}
\theta = 2 \cos^{-1}(w), \quad
\vec{u} = \frac{\vec{v}}{\lVert \vec{v} \rVert}, \quad
\vec{r} = \theta \cdot \vec{u},
\end{equation}
where $\vec{r}$ denotes the resulting Rodrigues vector.
When $\theta$ is close to zero, $\vec{v}$ also becomes small, allowing stable
approximation without numerical instability.

By representing joint rotations as Rodrigues vector sequences, each joint motion
is expressed as a consistent vector-valued temporal signal.
This unified formulation facilitates smooth optimization of motion transitions
across stitched motion segments using temporal Laplacian editing.

Apparent axis sign changes may be observed in Rodrigues vector visualizations due
to the redundancy of axis--angle representations, but these do not affect the
proposed temporal optimization.

\subsection{Smooth Motion Stitching via Laplacian Optimization}
\label{sec:laplacian_editing}

To achieve smooth stitching of heterogeneous motion segments, we formulate
motion editing as a temporal Laplacian optimization problem in Rodrigues vector
space, following the Laplacian deformation framework of
Sorkine et al.~\cite{Sorkine2004Laplacian}.
Temporal second-order differences of joint rotations are preserved to maintain
characteristic motion variations while suppressing discontinuities at segment
boundaries.

Let the Rodrigues vector time series of a motion be
$\mat{R} \in \mathbb{R}^{T \times 3}$.
Given an original motion $\mat{R}^{\mathrm{ori}}$ and a reference motion
$\mat{R}^{\mathrm{ref}}$, we first construct a target sequence
$\mat{R}^{\mathrm{tar}}$ by concatenating selected segments:
\begin{equation}
\mat{R}^{\mathrm{tar}}
=
\mat{R}^{\mathrm{ori}}[0{:}t_s)
\oplus
\mat{R}^{\mathrm{ref}}[t'_s{:}t'_e)
\oplus
\mat{R}^{\mathrm{ori}}[t_e{:}T)\, ,
\end{equation}
where $\oplus$ denotes temporal concatenation.

To avoid directly propagating numerical discontinuities caused by segment
replacement, the target Laplacian term is defined by concatenating Laplacians
from the original and reference motions:
\begin{equation}
\delta^{\mathrm{tar}}
=
\mat{L}\mat{R}^{\mathrm{ori}}[0{:}t_s)
\oplus
\mat{L}\mat{R}^{\mathrm{ref}}[t'_s{:}t'_e)
\oplus
\mat{L}\mat{R}^{\mathrm{ori}}[t_e{:}T)\, .
\end{equation}

The optimized motion $\mat{R}$ is obtained by minimizing the following energy:
\begin{equation}
\min_{\mat{R}}\;
\|\mat{R}-\mat{R}^{\mathrm{tar}}\|^2
+
\lambda \|\mat{L}\mat{R}-\delta^{\mathrm{tar}}\|^2 .
\label{eq:lap_energy}
\end{equation}

Since the objective is quadratic, the optimal solution is obtained by solving
the linear system
\begin{equation}
\left(
\mat{I}
+
\lambda \mat{L}^\top \mat{L}
\right)\mat{R}
=
\mat{R}^{\mathrm{tar}}
+
\lambda \mat{L}^\top \delta^{\mathrm{tar}} .
\label{eq:lap_linear}
\end{equation}
The resulting system matrix is symmetric and positive definite, allowing stable
and efficient optimization even for long motion sequences.

This formulation corrects rotational discontinuities introduced by segment
replacement while preserving the temporal characteristics of both the original
and reference motions.

\section{Experimental Setup}

We conducted two types of experiments to evaluate the proposed framework:
(1) motion stitching experiments to assess smooth integration of motion segments,
and (2) analysis of rotation axis inversion to examine the behavior of the
optimization under numerically discontinuous rotational representations.

All experiments were performed using standard motion capture data from Adobe
Mixamo\footnote{\url{https://www.mixamo.com}}.
Motion editing and visualization were carried out using a custom toolchain based
on \textbf{Blender} and Python, where joint rotations were extracted, converted
into Rodrigues vector space, and optimized using temporal Laplacian editing.

All optimization procedures were executed on a CPU using a Python
implementation.
For motion sequences of approximately 1,000 frames, the processing time was
around 0.1 seconds.
Although solving the Laplacian linear system dominates the computation, its cost
scales linearly with the number of frames, remaining practical for interactive
use.

\section{Motion Stitching Results}

\subsection{Smooth Motion Stitching via Laplacian Editing}

\begin{figure}[htbp]
\centering
\includegraphics[width=0.9\linewidth]{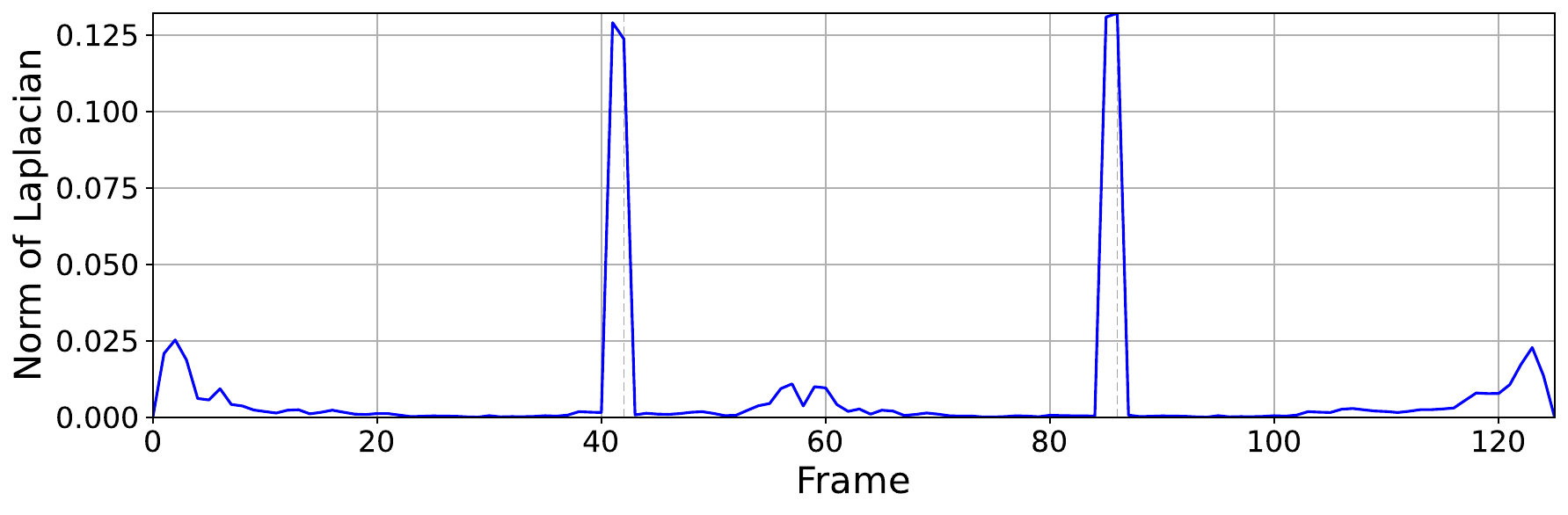}
\includegraphics[width=0.9\linewidth]{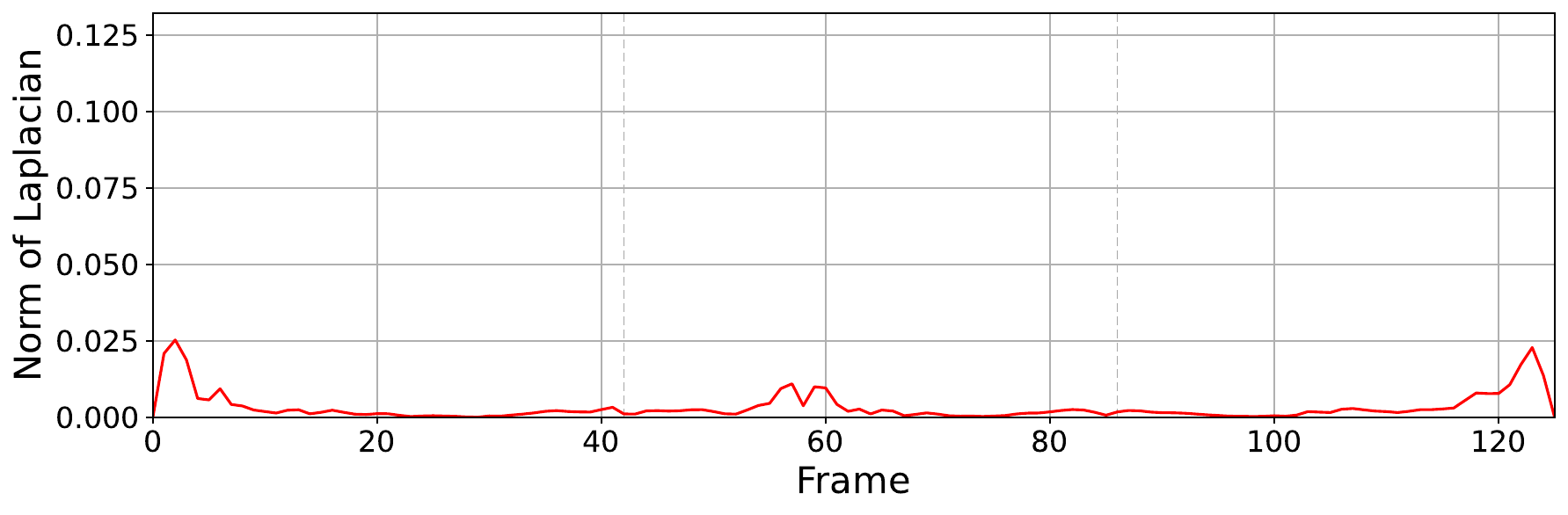}
\caption{Edited punch motions without (top) and with (bottom) Laplacian editing,
showing the temporal norm of Laplacian components of Rodrigues vectors
for a representative joint.}
\label{fig:laplacian-effect}
\end{figure}

We evaluate the effect of Laplacian editing on motion stitching by comparing
edited results obtained without and with Laplacian optimization, as shown in
\fref{fig:laplacian-effect}.

Without Laplacian editing, direct cut-and-paste operations introduce abrupt
rotational changes and unstable artifacts near the stitching boundaries.
In contrast, Laplacian optimization enforces smooth temporal transitions of
Rodrigues vectors, leading to coherent and visually natural motion.

These results indicate that Laplacian editing plays a critical role in achieving
smooth motion stitching beyond naive concatenation.
The corresponding animation results are included in the supplemental video.

\subsection{Motion Stitching and Comparison with Standard Quaternion Interpolation}

The supplemental video presents representative examples of cross-category
motion stitching using the proposed method, including a transition from walking
to jumping jacks.

We compare the proposed method with standard quaternion-based interpolation as
implemented in Blender’s built-in animation system.
Interpolation methods often introduce visible artifacts near transition
boundaries, and increasing the transition window tends to average out
characteristic motion dynamics.
In contrast, Laplacian editing enforces temporal consistency through global
optimization, enabling smooth transitions while preserving motion-specific
features without requiring explicit blending intervals.

The effect of blend width is further examined using punch motions with different
intensities.
For moderate differences, the proposed method produces stable transitions
without explicit tuning, whereas quaternion interpolation requires careful
adjustment.
For larger differences, interpolation results in ambiguous blending,
while Laplacian editing maintains coherent motion structure.

These results demonstrate that the proposed framework provides robust motion
stitching without reliance on manually specified blending regions.
The corresponding animations are included in the supplemental video.

\subsection{Seamless Integration of Cross-Category Motions}

We further demonstrate seamless integration of cross-category motion segments
using a composite example.
A walking motion is used as a base sequence, into which jumping and kicking
segments are inserted.

\begin{figure*}[tb]
  \centering
  \includegraphics[width=1.0\linewidth]{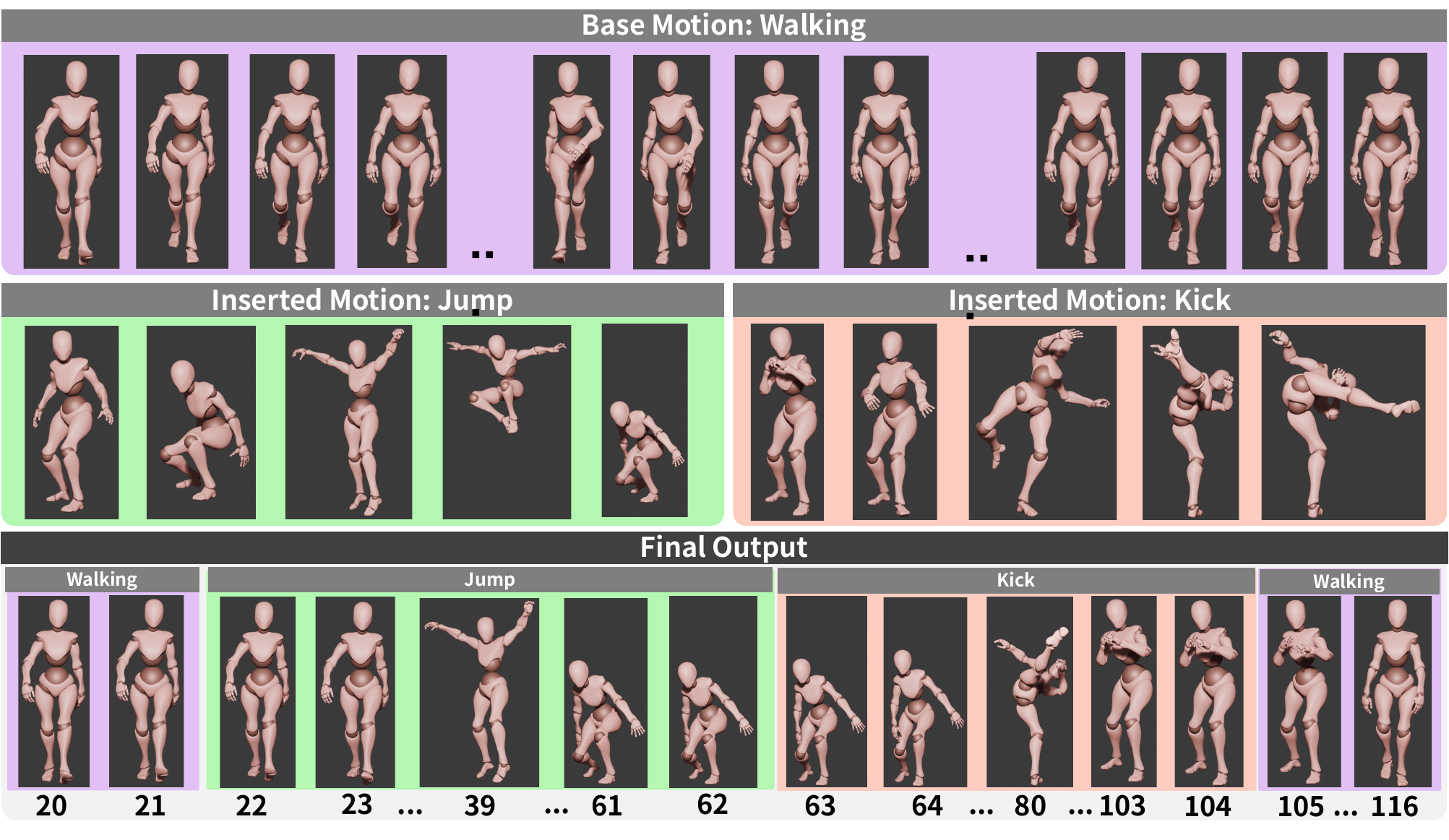}
  \caption{
Seamless integration of cross-category motions.
Top: base motion (walking).
Middle: inserted segments (jumping and kicking).
Bottom: integrated motion generated by the proposed method, showing smooth
transitions via Laplacian editing.
  }
  \label{fig:fusion_walk_jump_kick_seq}
\end{figure*}

As shown in \fref{fig:fusion_walk_jump_kick_seq}, simple cut-and-paste editing
would introduce noticeable rotational discontinuities at the insertion
boundaries.
In contrast, Laplacian editing corrects these discontinuities and produces
smooth transitions that remain visually coherent with the surrounding motion.

This example shows that the proposed framework can integrate motion segments
from different motion categories into a unified sequence, supporting flexible
motion editing and composition.
The corresponding animation results are included in the supplemental video.

\subsection{Locomotion Stitching via Root Translation Optimization}

Although our formulation focuses on rotational motion in Rodrigues vector space,
the same Laplacian framework can be applied to root translations.
By optimizing relative root displacements along the temporal axis,
continuous locomotion can be preserved even for non-in-place motions. As demonstrated in the supplemental video,
forward transitions such as running or walking into dynamic actions
maintain coherent global movement while faithfully reproducing
the original relative motion characteristics.

\section{Discontinuity Analysis of Rodrigues Vectors}

Rotational discontinuity is evaluated using the rotation magnitude $\theta$ and
the inner product of rotation axes between consecutive frames,
\begin{align}
\vec{u}_t \cdot \vec{u}_{t+1}
=
\frac{\vec{r}_t}{\|\vec{r}_t\|}
\cdot
\frac{\vec{r}_{t+1}}{\|\vec{r}_{t+1}\|},
\end{align}
where $\vec{r}_t$ denotes the Rodrigues vector at frame $t$.
For sufficiently large rotations, inner product values close to $1$ indicate
rotational continuity, while lower values imply axis misalignment and potential discontinuities.

\subsection{Rotational Continuity within Individual Motions}

\begin{figure}[htbp]
    \centering
    \includegraphics[width=1\linewidth]{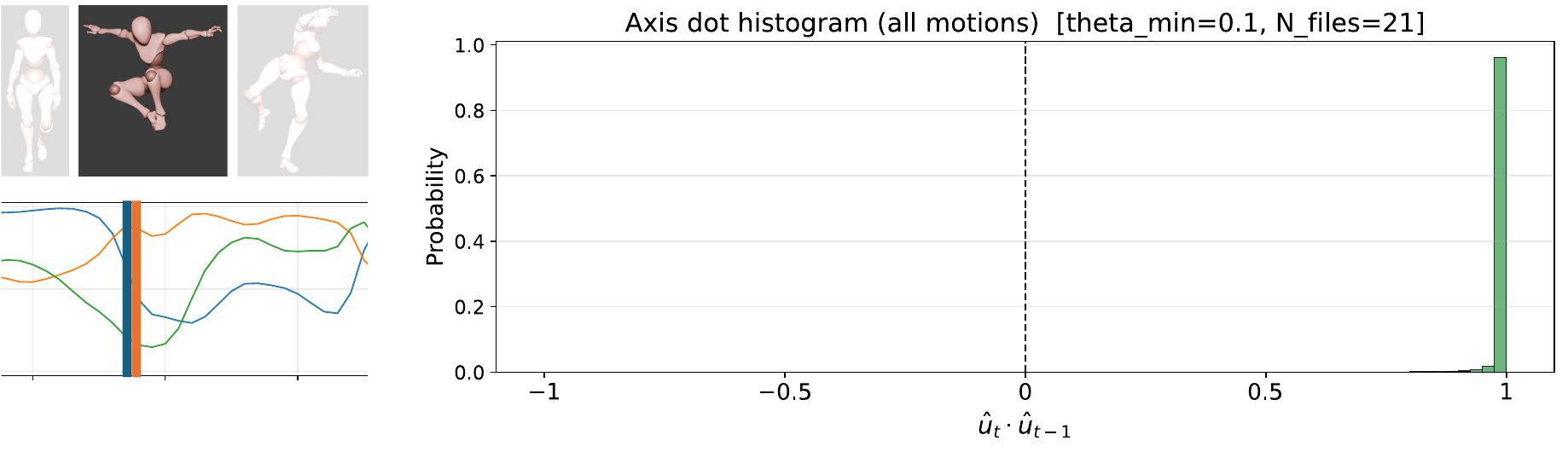}
    
    \caption{Histogram of rotation-axis inner products $\vec{u}_t \cdot \vec{u}_{t+1}$ aggregated over all motions, showing strong rotational continuity within individual motions.}
    \label{fig:hist_axis_dot_all_motions}
\end{figure}

We analyze rotational continuity within individual motions using multiple
motions from Adobe Mixamo.
Joint rotations are converted to Rodrigues vectors, and the inner product of
rotation axes between consecutive frames,
$\vec{u}_t \cdot \vec{u}_{t+1}$, is computed for all joints and frames.
The aggregated values are visualized as a histogram
(\fref{fig:hist_axis_dot_all_motions}).

The results show a strong concentration of values near $1$, indicating that
rotational discontinuities are rare within individual motions.
This reflects the smooth temporal evolution of joint rotations and rotation
axes in the Rodrigues representation.

\subsection{Rotational Continuity at Motion Stitching Points}

\begin{figure}[htbp]
    \centering
    \includegraphics[width=1\linewidth]{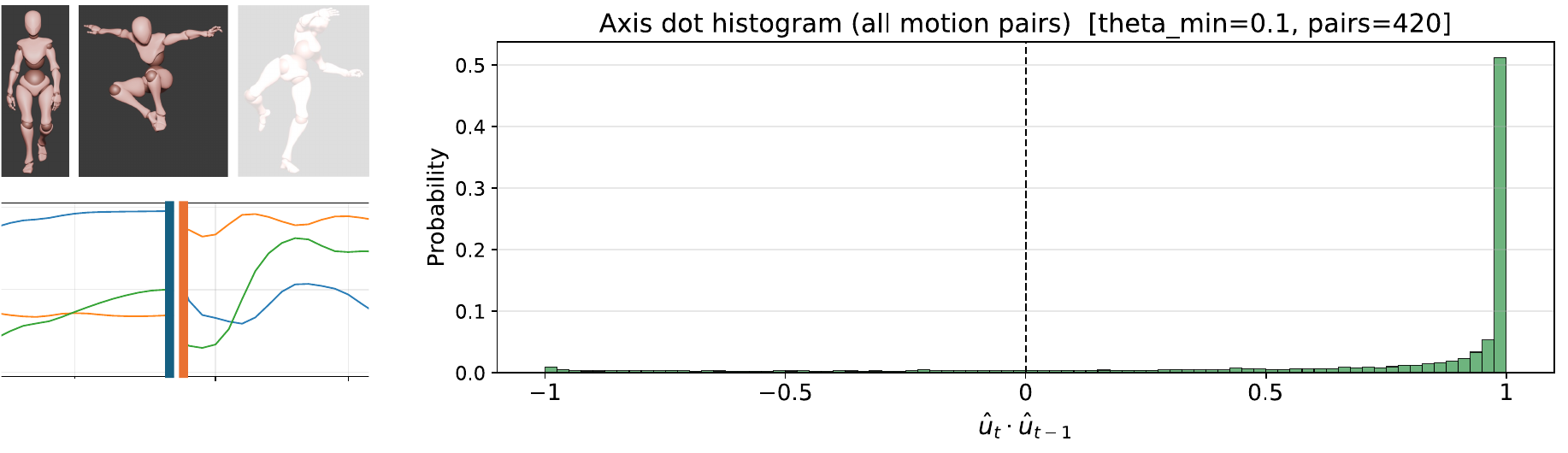}
    
    \caption{Histogram of rotation-axis inner products
             $\vec{u}_t \cdot \vec{u}_{t+1}$ across motion boundaries.
             $77.6\%$ of the values exceed $0.5$.}
    \label{fig:hist_axis_dot_all_motion_pairs}
\end{figure}

We evaluate rotational continuity at motion stitching points by sampling motion
pairs and analyzing frame pairs across non-contiguous boundaries.
For each joint, the inner product of rotation axes,
$\vec{u}_t \cdot \vec{u}_{t+1}$, is computed before and after the stitching point
and aggregated as a histogram
(\fref{fig:hist_axis_dot_all_motion_pairs}).

Although the values are generally concentrated near $1$, the distribution shows
a broader spread compared to that within individual motions.
In total, $77.6\%$ of samples exceed $0.5$, indicating that rotations remain
largely continuous in most cases, while axis mismatches can still occur at a
subset of stitching points.

\subsection{Intentional Axis Flipping: Synthetic Test}

We examine the behavior of Laplacian optimization under numerical
discontinuities introduced by intentional axis inversion in the Rodrigues
vector representation.

Given a Rodrigues vector $\vec{r}=\theta\vec{u}$, an equivalent axis-inverted
representation is defined as
\begin{equation}
\vec{r}^{\mathrm{inv}} = (2\pi - \theta)(-\vec{u}) .
\end{equation}
Although $\vec{r}$ and $\vec{r}^{\mathrm{inv}}$ describe the same rotation,
their numerical values differ significantly, introducing artificial
discontinuities in temporal sequences.

In the experiment, axis inversion is applied only to the replacement segment,
yielding a sequence that is rotationally continuous but numerically
discontinuous.
The resulting stitched motions are evaluated visually using the examples shown in \fref{fig:sign_flip_merged_compare}(a,b) and are also included in the supplemental video.

Without axis inversion, Laplacian optimization produces smooth transitions
across stitching boundaries.
In contrast, intentional axis inversion destabilizes the optimization and leads
to visible artifacts, demonstrating the sensitivity of Laplacian editing to
numerical discontinuities.

\begin{figure}[htbp]
    \centering

    \begin{subfigure}{\linewidth}
        \centering
        \includegraphics[width=\linewidth]{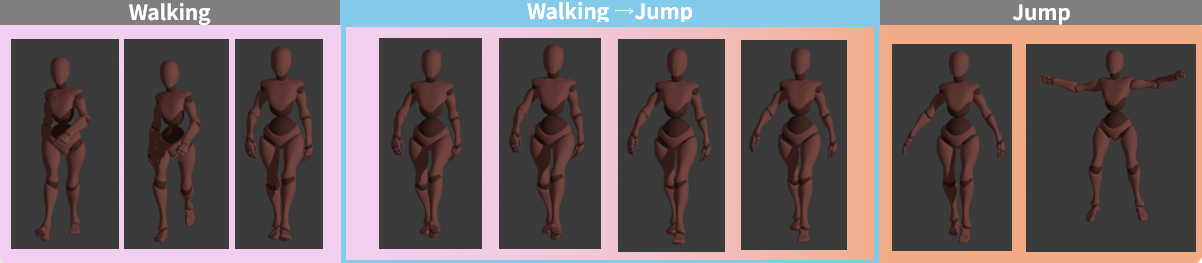}
        \caption{Stitched motion without axis flipping}
        \label{fig:sign_flip_merged}
    \end{subfigure}

    \vspace{2mm}

    \begin{subfigure}{\linewidth}
        \centering
        \includegraphics[width=\linewidth]{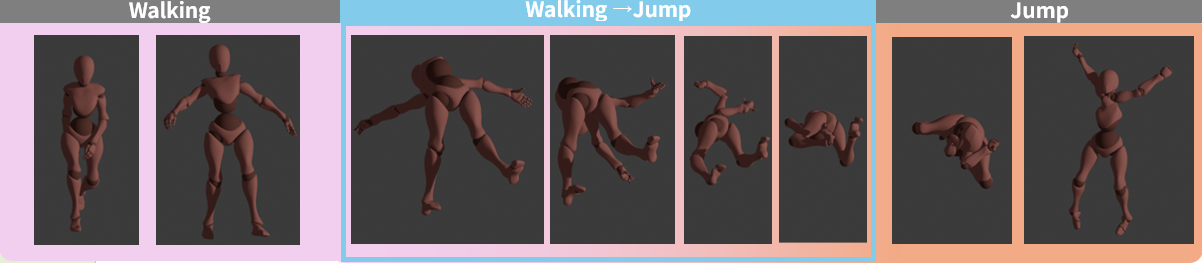}
        \caption{Stitched motion with intentional axis flipping
                 in the replaced segment}
        \label{fig:sign_flip_merged_flip}
    \end{subfigure}

    \caption{Effect of intentional axis flipping on Laplacian-based motion stitching.}
    \label{fig:sign_flip_merged_compare}
\end{figure}

\subsection{Stitching Results for Detected Axis Flip Cases}

From the analysis of rotation-axis inner products across all motion pairs
(\fref{fig:hist_axis_dot_all_motion_pairs}), a small number of stitching cases
with strongly negative values were detected, indicating potential axis inversion.
Among these, cases satisfying $\vec{u}_t \cdot \vec{u}_{t+1} < -0.9$ were selected
for further evaluation.

We applied motion stitching followed by Laplacian optimization to five such
motion pairs extracted from real motion data.
In all tested cases, the motions were smoothly connected without visible
artifacts, in contrast to the unstable behavior observed in the intentional
axis-flipping synthetic test.
These examples are shown in the supplemental video.

\begin{figure}[htbp]
    \centering
    \includegraphics[width=0.9\linewidth]{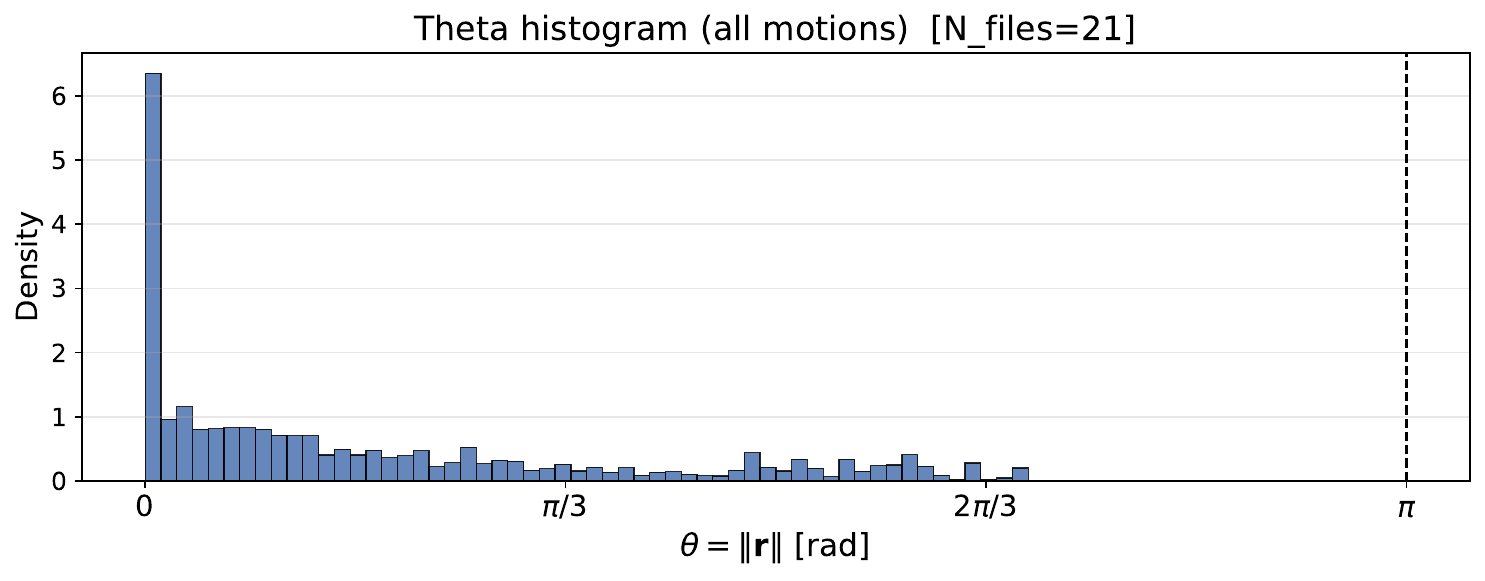}
    
    \caption{Distribution of rotation angles $\theta$ in real motion data.
The values remain within $[0,\pi]$, indicating numerical stability in practical motion stitching.}
    \label{fig:hist_theta_all_motions}
\end{figure}

This difference can be explained by the rotation angle $\theta$.
In real motion data, rotation angles are confined to $[0,\pi]$
(\fref{fig:hist_theta_all_motions}), whereas intentional axis inversion introduces
rotations of the form $2\pi-\theta$, leading to large numerical discontinuities.

These results indicate that artifacts in motion stitching arise primarily from
discontinuities in rotation magnitude rather than axis inversion itself.
Consequently, detected axis flip cases in real motion data do not pose a
significant practical problem for Laplacian-based motion stitching.

\section{Conclusion and Future Work}

We presented a motion editing framework for smooth motion stitching
based on Laplacian optimization in Rodrigues vector space.
By representing joint rotations as continuous vectors,
the method enables stable temporal smoothing
while suppressing cut-and-paste discontinuities.

Experiments demonstrate robust motion stitching
without learning-based models or complex parameter tuning.
Our analysis further clarifies why synthetic axis-flipping
instabilities do not arise in real motion data,
emphasizing the importance of rotational representation
for coherent motion transitions.

Future work includes automatic reference selection,
adaptive parameter control, semantics-aware motion editing,
and user-defined root motion constraints for controllable locomotion.
Extending the framework beyond joint-level kinematics,
the integration of electromyography (EMG) signals reflecting muscle activation
may further connect internal motor intent with observed motion dynamics.

\bibliography{main}
\bibliographystyle{IEEEtran} 

\end{document}